\documentclass[sigconf]{acmart}

\usepackage{booktabs} 
\usepackage{listings}
\usepackage{tabularx}

\usepackage{url}

\usepackage{subfig}

\usepackage{natbib}

\newcommand{\rev}[1]{\textcolor{black}{#1}}

\setcopyright{rightsretained}

\acmDOI{}

\acmISBN{}

\acmConference[PCSC2023]{Philippine Computing Science Congress}{March 2023}{City of Cebu, Philippines}
\acmYear{2023}
\copyrightyear{2023}

\acmArticle{}
\acmPrice{}

\editor{}
\editor{}
\editor{}

\begin{document}
\title[Transforming Chatbot Dialogues into Data Mart Schema for Visualization]{From Unstructured to Structured: Transforming Chatbot Dialogues into Data Mart Schema for Visualization}

\author{Mark Edward M. Gonzales}
\orcid{0000-0001-5050-3157}
\affiliation{%
  \institution{De La Salle University}
  \streetaddress{Taft Avenue, Malate}
  \city{Manila}
  \country{Philippines}
  \postcode{1004}
}
\email{mark\_gonzales@dlsu.edu.ph}

\author{Elyssia Barrie H. Ong}
\affiliation{%
  \institution{De La Salle University}
  \streetaddress{Taft Avenue, Malate}
  \city{Manila}
  \country{Philippines}
  \postcode{1004}
}
\email{elyssia\_ong@dlsu.edu.ph}

\author{Charibeth K. Cheng}
\affiliation{%
  \institution{De La Salle University}
  \streetaddress{Taft Avenue, Malate}
  \city{Manila}
  \country{Philippines}
  \postcode{1004}
}
\email{charibeth.cheng@dlsu.edu.ph}

\author{Ethel Chua Joy Ong}
\affiliation{%
  \institution{De La Salle University}
  \streetaddress{Taft Avenue, Malate}
  \city{Manila}
  \country{Philippines}
  \postcode{1004}
}
\email{ethel.ong@dlsu.edu.ph}

\author{Judith J. Azcarraga}
\affiliation{%
  \institution{De La Salle University}
  \streetaddress{Taft Avenue, Malate}
  \city{Manila}
  \country{Philippines}
  \postcode{1004}
}
\email{judith.azcarraga@dlsu.edu.ph}

\renewcommand{\shortauthors}{M.E.M. Gonzales, E.B.H. Ong, C.K. Cheng, E.C.J. Ong \& J.J. Azcarraga}

\begin{abstract}
Schools are among the primary avenues for public healthcare interventions. With resource limitations posing challenges to the routine conduct of health and wellness checks in Philippine public schools, the deployment of a chatbot-assisted health monitoring system may provide an alternative method. However, deriving insights from raw conversations is not straightforward due to the expressiveness of natural language that causes variances in the input. In this paper, we present a \rev{process} for transforming unstructured dialogues into a structured schema. The \rev{process} comprises four stages: \textit{(i)} processing the dialogues through entity extraction and data aggregation, \textit{(ii)} storing them as NoSQL documents on the cloud, \textit{(iii)} transforming them into a star schema for online analytical processing and building an extract-transform-load workflow, and \textit{(iv)} creating a web-based dashboard for visualizing summarized data and reports. Performance evaluation of this dashboard showed that increasing the number of stored dialogues by a factor of \(10^5\) increased the loading time for the display of roll-up, drill-down, and filter results by around only one second.
\end{abstract}

%
%



\maketitle

\section{Introduction}

The promotion of health and wellness among primary school students is one of the thrusts of the United Nations Sustainable Development Goals \cite{52]un}. In the Philippines, schools are the foremost avenues for public health intervention \cite{43]ruff2022}, with national and regional programs, such as the Essential Heath Care Program \cite{35]monse2013}, \textit{Oplan Kalusugan sa Department of Education} \cite{42]oplan2019}, and free vaccination initiatives \cite{23]guadalquiver2018}, instituted to control the levels of preventable diseases among schoolchildren.

Despite these efforts, a significant barrier to performing routine wellness assessments in public elementary schools can be ascribed to a lack of manpower, as the usual approach to health monitoring requires nurses or doctors to facilitate the review of body systems. Moreover, student health records are typically stored as physical documents, which may become fragmented with the accumulation of more data over time and may be difficult to readily retrieve, update, and share across healthcare providers \cite{4]antonio2016, 51]thomas2009}.

The deployment of an automated health monitoring system may help in facilitating routine wellness checks without the presence of a medical practitioner and store gathered information in the cloud. Since children may not be accustomed to navigating the highly structured paths in traditional WIMP (windows, icons, menus, and pointers) interfaces, integrating a chatbot into this system has the potential of capitalizing on a chatbot’s personified and anthropomorphic features to provide a familiar way for users to more freely interact and express themselves \cite{54]xiao2020}. 

Aside from employing a chatbot with speech and text processing abilities to interpret audio input and formulate intelligent responses, the health monitoring system should extract insights from the children’s conversations with the chatbot, update their health information in the cloud, and provide analytics and data visualization for healthcare professionals. Actualizing these necessitates a \rev{process} that considers the unstructured nature of conversations, expressiveness of natural language input, and database design.
The \rev{process} includes transforming unstructured dialogues into a structured schema for data visualization. In particular, this \rev{process} attempts to provide a way to:

\begin{itemize}
    \item Process the dialogues by performing entity extraction and data aggregation;
    \item Store the processed dialogues in the cloud;
    \item Transform the stored dialogues into a schema that facilitates online analytical processing (OLAP) and build an automated extract-transform-load (ETL) workflow; and
    \item Create a web-based visualization dashboard that summarizes the data from the children’s conversations with the chatbot.
\end{itemize}

\section{Related Works}
Chatbots are computer programs that can interact with human users using natural language, thereby simulating coherent conversations \cite{1]alrazaq2019, 39]arrays}. Chatbots tailored for healthcare allow clinicians to easily organize digital health records, retrieve patient data, and provide information on observed side effects and drug interactions \cite{54]xiao2020}. Aside from assisting healthcare professionals, some are also designed for personal health monitoring and management \cite{9]bates2019}.

\subsection{Dialogue Processing and Management}
As chatbots are involved in speech-based communication, they implement action selection methods in dialogue management to process input statements and return appropriate responses \cite{11]burgan2016dialogue, 31]laranjo2018}, with healthcare chatbots typically employing medical entity recognition \cite{17]ghosh2018} and semantic similarity measures. Quro \cite{17]ghosh2018}, a personalized healthcare assistant, uses medical entity recognition to extract symptoms from user input, where medical entities represent any medical concept (e.g., sign, symptom, disease, or drug), and detect a semantic relationship between the said entities.

Recently, semantic similarity measures for dialogue management have been adopted as an alternative to traditional pattern-matching approaches \cite{2]adel2020,10]bhirud2019}, removing the need to create scripting patterns, which may be time-consuming and error-prone. Adeh \textit{et al.} \cite{2]adel2020} implemented an algorithm for measuring fuzzy sentence similarity that takes two phrases or short texts and provides a similarity measure of meaning between the two based on given syntactic and semantic elements. In comparison to an established similarity measure based on semantic networks and corpus statistics \cite{33]li2006}, the use of fuzzy sentence similarity has been found to improve rule-matching, although it struggles in processing phrases with the word \textit{not} since it causes the misfire of some rules, indicating that special considerations have to be made for words associated with negative meaning.

Fuzzy matching has been commonly used by healthcare chatbots to serve a range of purposes, from as simple as improving word recognition despite writing or typographical errors \cite{34]liu2018}, to working with webhooks and support vector machines to recognize medical entities, as with IBM Watson Assistant \cite{40]rahman2022}.

\subsection{Healthcare System Database Design} 
Healthcare systems manage large amounts of heterogeneous data. With the increase in digital healthcare data in particular, the volume and types of data become salient considerations. Dispersed pieces of information have to be collected from patients, prompting the need to extract and collate data into one repository \cite{12]carchiolo2015}.

Data marts and warehouses can be helpful in this context, as they are repositories comprised of different data sources, all organized under a single schema for use in generating insights that aid in strategic decision-making \cite{47]sani2012, 48]silberschatz2020}. Star schemas are often adopted in healthcare data management research \cite{6]appah2018multidimensional, 36]narra2015, 49]talib2021} due to their simplicity, leading to effective query handling, fast aggregations, as well as faster query execution times due to fewer join operations. As these schemas comprise a central fact table connected to multiple dimension tables, the resulting denormalization of dimension tables introduces redundancy and, consequently, increased storage requirements \cite{37]oliva2018}. 

However, this is outweighed by the relatively lower cost of storage \cite{36]narra2015} and users’ preference for faster query operations across large volumes of data \cite{6]appah2018multidimensional, 28]johnson2008}. These characteristics make star schema a logical design choice for more efficient execution of OLAP operations \cite{49]talib2021}. 

\subsection{Healthcare Data Visualization} 
Data visualization is a method of presenting information using visual encodings of quantitative data \cite{56]yang2020}, enabling decision-makers to interpret results and identify patterns more easily. As such, visualizations are widely used to deliver scientific information, including health data, to both professional and general audiences. 

With the large variety of available data visualizations, several works have looked into the format preferences of target audiences upon being presented with health-related data. Fortin \textit{et al.} \cite{16]fortin2001} investigated women’s visualization preferences to present the risk of heart disease, hip fracture, and breast cancer in women. They found that most respondents preferred simple bar charts to other traditional data visualizations. 

Similarly, Dolan and Iadarola \cite{15]dolan2008} investigated six types of visualizations for presenting information on cancer and its prevention: augmented bar charts, icon arrays, flowcharts, and three other separate combinations of the aforementioned visualizations. It was found that, while bar charts and flowcharts were the preferred formats, presenting a combination of data visualization types is preferable to presenting a single standalone visualization.

The development of more creative and sophisticated data presentation formats has also prompted studies to look into the effect of interactivity on the understandability of visualizations, although studies that focus specifically on presenting healthcare-related data are limited \cite{56]yang2020}. Ancker, Chan, and Kukafka \cite{3]ancker2009} assessed the usability of interactive graphs using icon arrays to display risk information. Their results indicated that the interactive graphs have the potential to present the magnitude of risk, as well as the feeling associated with it. This is in consonance with studies positing that high levels of interactivity in computer-mediated communication may positively influence users’ processing of information, attitude towards health, and emotional investment in health-related messages \cite{56]yang2020}.

\section{Design and Implementation}
\begin{figure}[!t]
    \includegraphics[width = \linewidth]{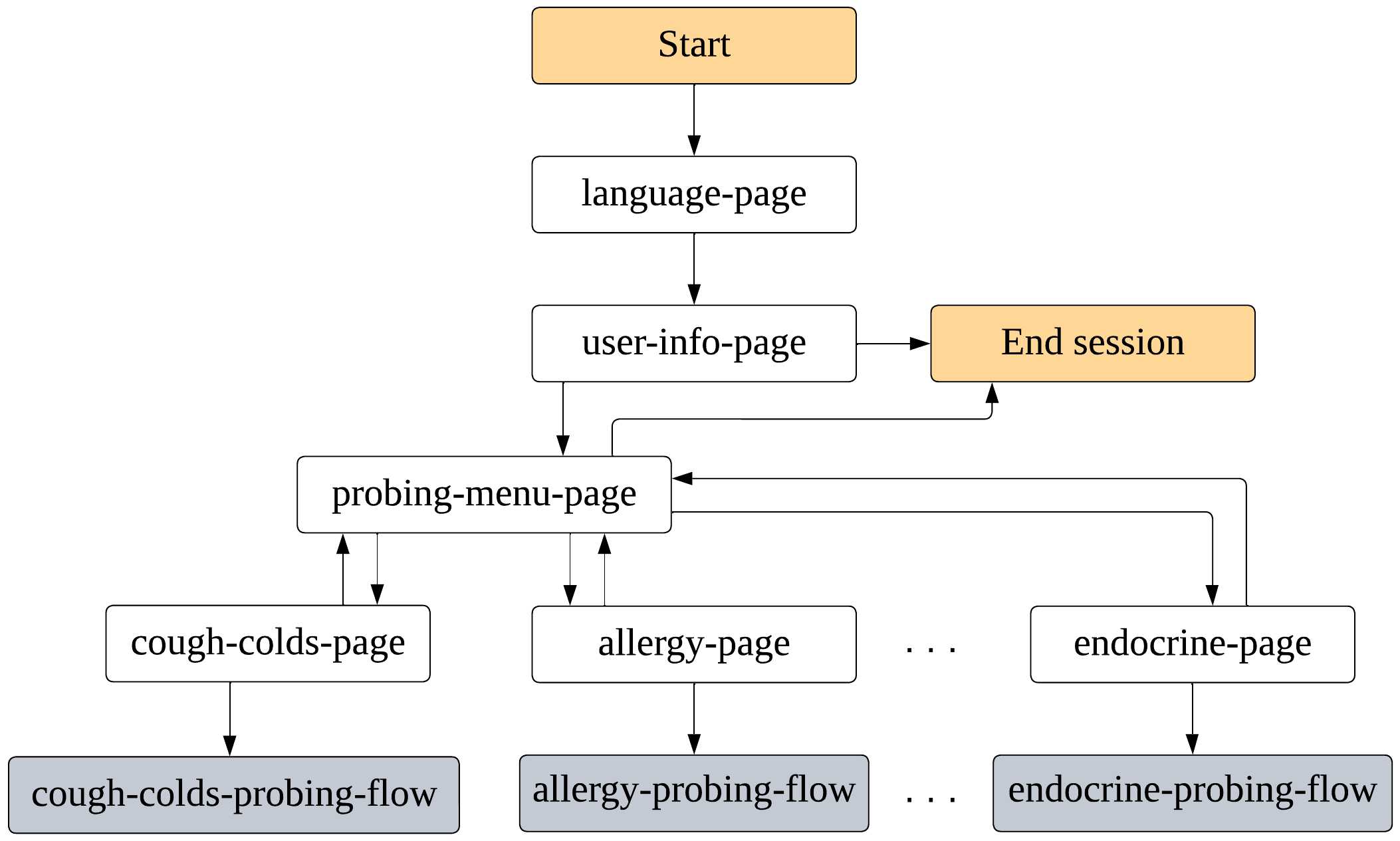}
    \caption{Modularization of conversation flow. \textmd{A session with the chatbot starts by asking for the child’s preferred language and information (username and data privacy consent). If the child does not give their consent, the session terminates. Otherwise, the chatbot proceeds to a review of body systems, which is handled under the hood by multiple modules (gray).}}
    \label{modularization}
\end{figure}

\begin{figure}[!t]
    \includegraphics[width = \linewidth]{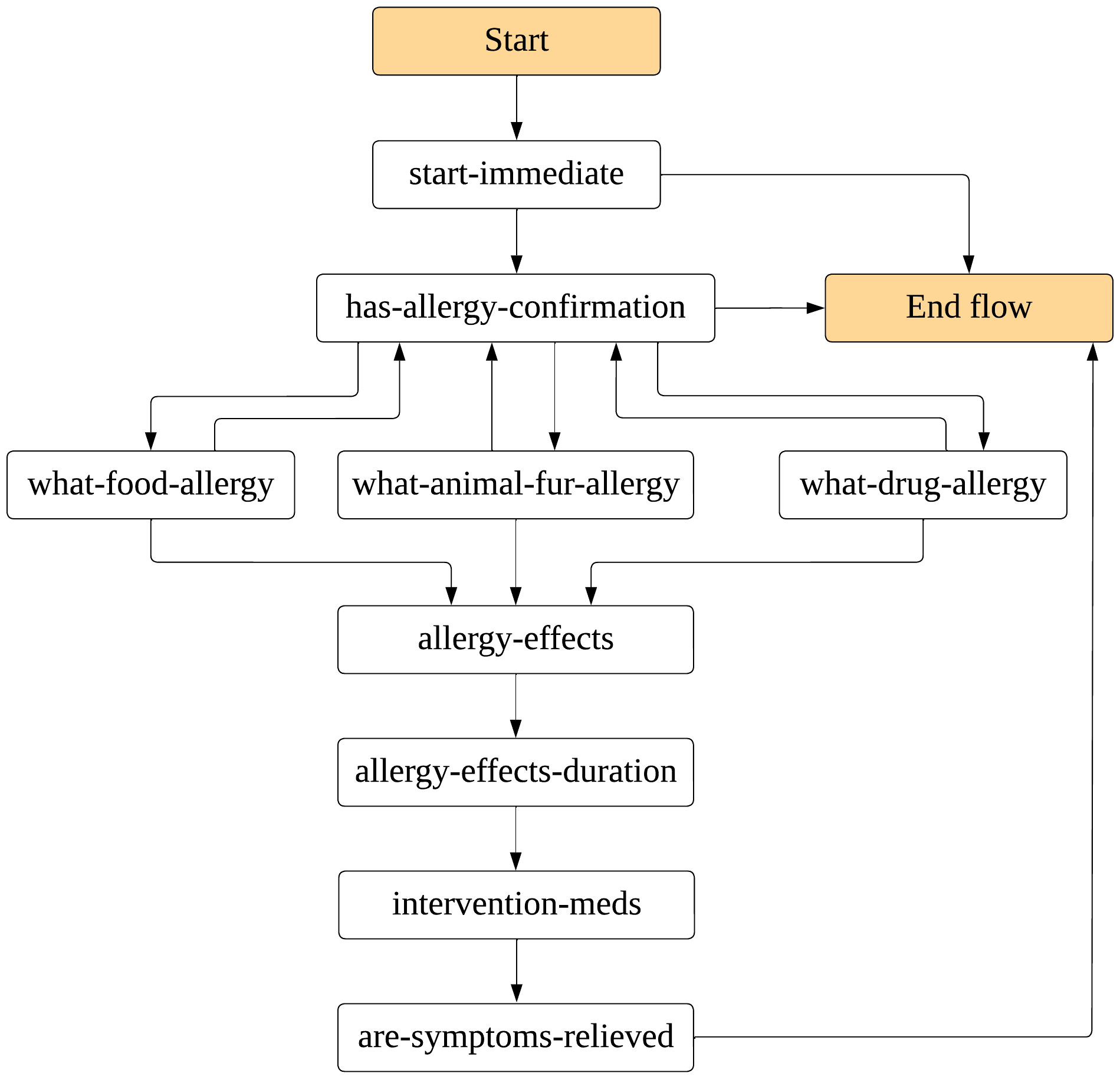}
    \caption{Conversation flow for the allergy probing module. \textmd{DialogFlowCX \cite{20]dialogflowcx} models each flow as a finite-state machine, with the pages as the states and the state handlers as the transitions. The intermediate page  after the start of the flow sets the parameters for the communication between the chatbot and the knowledge base.}}
    \label{conv-flow}
\end{figure}

The chatbot was created using DialogFlowCX \cite{20]dialogflowcx}, a cloud-based service launched by Google in 2021 for designing conversational agents. Figure~\ref{modularization} illustrates the modularization of the conversation flow into smaller conversation flows corresponding to the sections of an adapted version of the Pediatric Review of Systems by Heartland Community Health Center \cite{25]heartland2012}. Each of these smaller conversation flows is, in turn, composed of several connected pages containing questions patterned after the Pediatric Review of Systems; an example is illustrated in Figure~\ref{conv-flow} for probing allergy symptoms. 

\begin{figure*}[!t]
    \includegraphics[width = \linewidth]{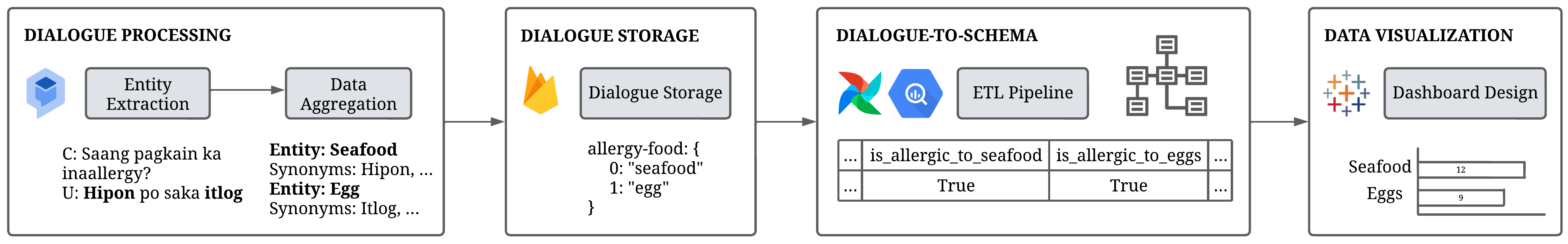}
    \caption{Block diagram of the \rev{process}. \textmd{The \rev{process} consists of four stages: \textit{(i)} processing the dialogues using DialogFlowCX, \textit{(ii)} storing them in Cloud Firestore, \textit{(iii)} transforming them to a star schema, with the database stored in BigQuery and an ETL workflow built and running in Apache Airflow, and \textit{(iv)} creating a Tableau dashboard for data visualization.}}
    \label{pipeline}
\end{figure*}

The remainder of this section presents each of the \rev{stages in the process of} transforming unstructured dialogues with this chatbot into a structured schema for data visualization. Figure~\ref{pipeline} \rev{presents the block diagram, from processing and storing the dialogues to transforming them into a data mart schema and finally creating a dashboard for data visualization}.

\subsection{Dialogue Processing}
The first stage in the \rev{process} is processing the dialogues. Since the bot-child conversations include open-ended questions  (e.g., \textit{Saang pagkain ka inaallergy?}), it is imperative to take the expressiveness of natural language input into account.

\subsubsection{Expressiveness of Natural Language}
The dialogue processing component should be capable of supporting two fundamental natural language understanding (NLU) functions: entity extraction and data aggregation. Entity extraction involves the recognition of keywords or phrases from the child’s utterance in the context of their intent. A sample conversation \rev{between the chatbot and the child} is given in Listing~\ref{dialogue}; the entities are in bold.

\lstdefinestyle{interfaces}{
  float=tp,
  floatplacement=tbp,
  abovecaptionskip=-5pt
}

\lstset{emph = {Meron, Hipon, Wala, Hirap, huminga, nagpapantal, ointment, itlog}, emphstyle = \textbf, columns=fullflexible, keepspaces=true, breaklines=true, frame=tb, gobble=4, breakindent=0pt, label={regex}, belowcaptionskip=14pt}
\begin{lstlisting}[style = interfaces, label = dialogue, caption = {
    Sample dialogue with entity extraction. \textmd{\textit{C} refers to the chatbot, while \textit{U} refers to the end user (child). The entities extracted in the context of the end user's intent are in bold.}}]
    C: May allergy ka ba sa pagkain?
    U: Meron
    C: Saang pagkain ka inaallergy?
    U: Hipon saka po itlog
    C: May allergy ka ba sa balahibo ng hayop?
    U: Wala po
    C: Anong nararamdaman mo kapag inaallergy ka?
    U: Hirap huminga saka po nagpapantal
    C: Anong gamot ang inilalagay o iniinom mo para sa allergy?
    U: May nilalagay si Mommy na ointment
\end{lstlisting}

Data aggregation involves grouping distinct words or phrases that are considered equivalent in the context of the conversation and assigning a single reference value for them. Table~\ref{word-rel} enumerates examples of categories of word relationships that have to be subjected to aggregation; the reference value for each group of words is in bold.

\begin{table}[t!]
  \caption{Word relationships subject to data aggregation. \textmd{The reference value for each group of words is in bold.}}
  \begin{tabularx}{\linewidth}{Xl}
    \toprule
    Relationship & Words \\
    \midrule
    Orthographic Variations & \textbf{tiyan}, tyan \\
    Misspellings & \textbf{penicillin}, penicilin, penisilin \\
    Inflections & \textbf{egg}, eggs \\
    Synonyms & \textbf{yes}, yeah, yep, of course \\
    Hypernyms/Hyponyms & \textbf{seafood}, crab, mussels, shellfish \\
    Named Entities & \textbf{vitamins}, Tiki Tiki, Poten-Cee \\
    Translations & \textbf{abrasion}, gasgas \\
    Elliptical Phrases \newline \hspace*{3mm} (Context: Age) & \textbf{3}, 3 year old, 3 years old \\
    Contextual Synonyms \newline \hspace*{3mm} (Context: Urine color) & \textbf{clear}, no color, white \\
  \bottomrule
\end{tabularx}
  \label{word-rel}
\end{table}

\subsubsection{Answer Integrator} \label{ans-int}
To perform entity extraction and data aggregation, an answer integrator was developed using the Entities feature \cite{21]entities} built into DialogFlowCX. A list of entity entries and equivalent words (termed \textit{entity synonyms} in DialogFlowCX) for each entity type was manually constructed to cover the word relationships in Table~\ref{word-rel}. Table~\ref{entities-allergy} provides an example for the entity type about food allergies, \rev{which include reference values such as \textit{nuts}, \textit{dairy}, \textit{egg}, and \textit{seafood}}.

\begin{table}[t!]
  \caption{Entity entries for the food allergies entity type}
  \begin{tabularx}{\linewidth}{lX}
    \toprule
    Reference Value & Selected Entity Synonyms \\
    \midrule
    nuts & nuts, peanut, mani \\
    dairy & dairy, cheese, yogurt, ice cream, milk, keso, queso, gatas \\
    egg & egg, itlog \\
    seafood & seafood, crab, mussels, shellfish, shrimp, pagkaing dagat, alimago, tahong, hipon \\
  \bottomrule
\end{tabularx}
  \label{entities-allergy}
\end{table}

The entity entry and synonym lists were primarily based on the conversations recorded from 227 respondents aged 6 to 7 years old who underwent a pediatric review of systems facilitated by partner nurses. They were subsequently expanded by the nurses serving as domain experts. 

The answer integrator performs fuzzy matching with the entity entries to extract the entities from the child’s utterance. To this end, the fuzzy matching functionality \cite{22]entities} of DialogFlowCX is invoked. Data aggregation is then performed by mapping the extracted entities to their reference values. If the fuzzy matching returns an empty set, a \texttt{sys.no-match-default event} is triggered, and the chatbot reprompts the child.

As similarly demonstrated in the work of Liu and Sundar \cite{34]liu2018}, and in HealthConsultantBot \cite{40]rahman2022}, the advantage of employing fuzzy matching over exact or regular expression-based matching is that misspellings, orthographic variations, and inflections can be accommodated to an extent without the need to manually add them to the entity synonym set or create hard-coded regular expression patterns that may become unwieldy or inadequate as the chatbot’s complexity increases. 

This approach is also capable of handling multiple-word entities, which is especially important since the typical Tagalog and Bisaya sentence structure places a pronoun subject between the negating particle and the verb. For example, \textit{hindi ko po alam} and \textit{wala ko kabalo} are fuzzily matched to \textit{hindi alam} and \textit{wala kabalo}, which both map to the reference value \textit{don’t know}.

However, it may yield false matches for words with a small edit distance; for instance, \textit{hipon} (\textit{shrimp}) may be matched with \textit{sipon} (\textit{colds}) if the latter is already the closest approximate match from among the entities. To mitigate this, orthographically similar words that map to distinct reference values were manually added to the entity entry and synonym lists. Fuzzy matching was also turned off for responses that should match with exact values (e.g., usernames). 

\subsection{Dialogue Storage}
The second stage in the \rev{process} is saving each processed dialogue as a JSON document and storing it in a collection hosted in Cloud Firestore \cite{19]cloudfirestore}, a scalable NoSQL database that provides seamless integration with other Google Cloud Platform services (including DialogFlowCX). A field-value pair in this document corresponds to a question and the entity reference value capturing the child’s response as determined by the answer integrator.

Since a dialogue is essentially a series of questions and responses, modeling it as a NoSQL document composed of field-value pairs allows for a straightforward and flexible representation compared to the table-based approach of relational databases \cite{48]silberschatz2020}. Moreover, some fields (e.g., food allergies and symptoms felt during an allergic reaction) should be capable of holding multiple values. While this is not an issue for JSON documents, support for multi-valued data types in relational databases varies depending on the database management system (DBMS) \cite{39]arrays}.

Table~\ref{doc-rep} shows the relevant field-value pairs in the document representation of the sample dialogue in Listing~\ref{dialogue}; fields storing the user and session information are also included. Since each document corresponds to a session with the chatbot, the document ID generated by Cloud Firestore also serves as the session ID.

\begin{table}[t!]
  \caption{Document representation of the sample dialogue in Listing~\ref{dialogue}}
  \begin{tabularx}{\linewidth}{lX}
    \toprule
    Field & Value \\
    \midrule
    \texttt{session\_id} & 123456789 \\
    \texttt{username} & juan-dela-cruz \\
    \texttt{sex} & M \\
    \texttt{data\_privacy\_consent} & True \\
    \texttt{allergy\_food} & seafood, egg \\
    \texttt{allergy\_animal\_fur} & None \\
    \texttt{allergy\_felt} & difficulty breathing, rashes \\
    \texttt{allergy\_intervention} & ointment \\
  \bottomrule
\end{tabularx}
  \label{doc-rep}
\end{table}

\subsection{Dialogue-to-Schema Transformation}
The third stage in the \rev{process} is transforming the stored dialogues into a structured data mart schema for OLAP. In order to take advantage of the suitability of relational databases and the expressive power of SQL for executing OLAP operations, a star schema and an ETL pipeline were designed.

\subsubsection{Star Schema Design} \label{star-schema-sec} 
\begin{figure}[!t]
    \includegraphics[width = \linewidth]{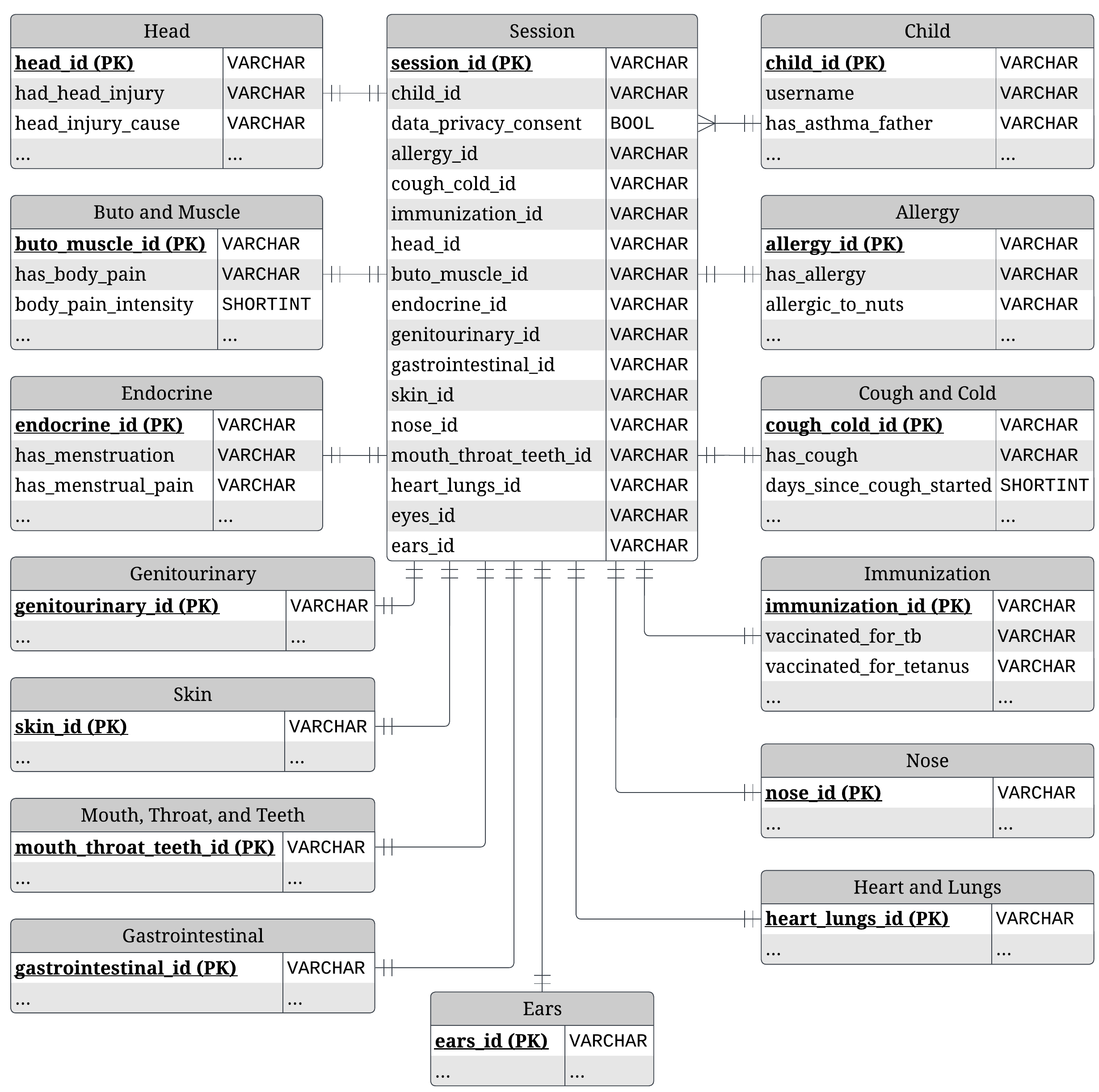}
    \caption{Star schema for transforming processed dialogues for OLAP. \textmd{The rationale for the decomposition of the collection in Cloud Firestore into these tables is to maintain congruence with the modularization of the conversation flow in DialogFlowCX. The use of a variable-length string (\texttt{VARCHAR}) as the data type for most columns is for accommodating \textit{yes}, \textit{no}, and \textit{don’t know} answers.}
}
    \label{star-schema}
\end{figure}

The database design follows a star schema. As shown in the entity relationship diagram in Figure~\ref{star-schema}, it consists of \textit{(i)} one fact table storing the session ID and the keys connecting it to the dimension tables and \textit{(ii)} 14 dimension tables corresponding to the sections in the adapted Pediatric Review of Systems. The primary key of the fact table is the session ID, while the primary key of each dimension table is a surrogate key generated during row insertion. 

The advantage of following a star schema is that dimension tables are denormalized, resulting in fewer tables \cite{28]johnson2008}. Consequently, the number of join statements needed to perform OLAP operations is reduced, simplifying queries and speeding up their execution \cite{48]silberschatz2020}. Due to these benefits, the star schema is implemented in several healthcare systems, as the database design makes the management of large amounts of heterogeneous data easier \cite{6]appah2018multidimensional, 36]narra2015, 37]oliva2018}. 

An important database design consideration is accommodating fields with multiple values. The approach taken in this schema is to split these values into multiple columns, which is possible since the answer integrator module (Section~\ref{ans-int}) enumerates all the entity reference values recognized by the chatbot. For illustration, Table~\ref{table-doc} shows \rev{some of} the relevant tables and fields (columns) in the transformation of the document in Table~\ref{doc-rep} following this schema; \rev{since the child and allergy information are stored in separate tables, surrogate keys (namely \texttt{child\_id} and \texttt{allergy\_id}) are assigned.}

\begin{table}[t!]
  \caption{Partial table representation of the document in Table~\ref{doc-rep}}
  \begin{tabularx}{\linewidth}{Xl}
    \toprule
    Field (Column) & Value \\
    \midrule
    Session Table & \\
    \midrule
    \texttt{session\_id} & 123456789 \\
    \texttt{child\_id} & 424242424 \\ 
    \texttt{allergy\_id} & 787878787 \\ 
    ... & ... \\
    \midrule 
    Child Table & \\
    \midrule 
    \texttt{child\_id} & 424242424 \\
    \texttt{username} & juan-dela-cruz \\
    ... & ... \\ 
    \midrule 
    Allergy Table & \\ 
    \midrule 
    \texttt{allergy\_id} & 787878787 \\ 
    \texttt{allergic\_to\_nuts} & no \\ 
    \texttt{allergic\_to\_eggs} & yes \\
    \texttt{allergic\_to\_seafood} & yes \\
    ... & ... \\ 
    \texttt{felt\_difficulty\_breathing} & yes \\ 
    \texttt{felt\_rashes} & yes \\ 
    ... & ... \\
    \texttt{intervention\_applied\_ointment} & yes \\ 
    \texttt{intervention\_away\_from\_allergens} & no \\
    ... & ... \\
  \bottomrule
  \end{tabularx}
  \label{table-doc}
\end{table}

While this approach entails adding a new column to the schema for every addition to the list of entity reference values, this was considered acceptable in the context of this project since the entity reference values (e.g., common food allergies and allergic reaction symptoms in medical literature or checklists \cite{7]australiansociety, 13]doh}) are not expected to change frequently.

Alternative options that were considered include storing multiple values using a multi-valued data type or as a comma-separated string. However, the former will introduce an additional layer of complexity in case of migration, as support for multi-valued data types is DBMS-dependent \cite{39]arrays}. Meanwhile, the latter is regarded as an antipattern since a comma-separated string has to be parsed before summary statistics can be extracted, thus making queries for OLAP aggregation more complex and inefficient \cite{30]karwin2010}. 

Designing a snowflake schema is another approach that was also considered. Although this offers scalability, it carries the overhead of creating two tables (i.e., a table to hold the possible values and a mapping table) for every multi-valued field. Enforcing this degree of normalization increases the number of join statements and, consequently, the execution time of OLAP queries \cite{48]silberschatz2020}.

Finally, although storing each value in a separate row may be adopted as a possible denormalization strategy, it comes with the caveat that a dialogue with the healthcare chatbot includes several multiple-response questions, thereby resulting in high levels of redundancy and significantly increased storage requirements. These may pose a challenge for the integration of the chatbot into a nationwide health monitoring system.

\subsubsection{ETL Pipeline}
\begin{figure}[!t]
    \includegraphics[width = \linewidth]{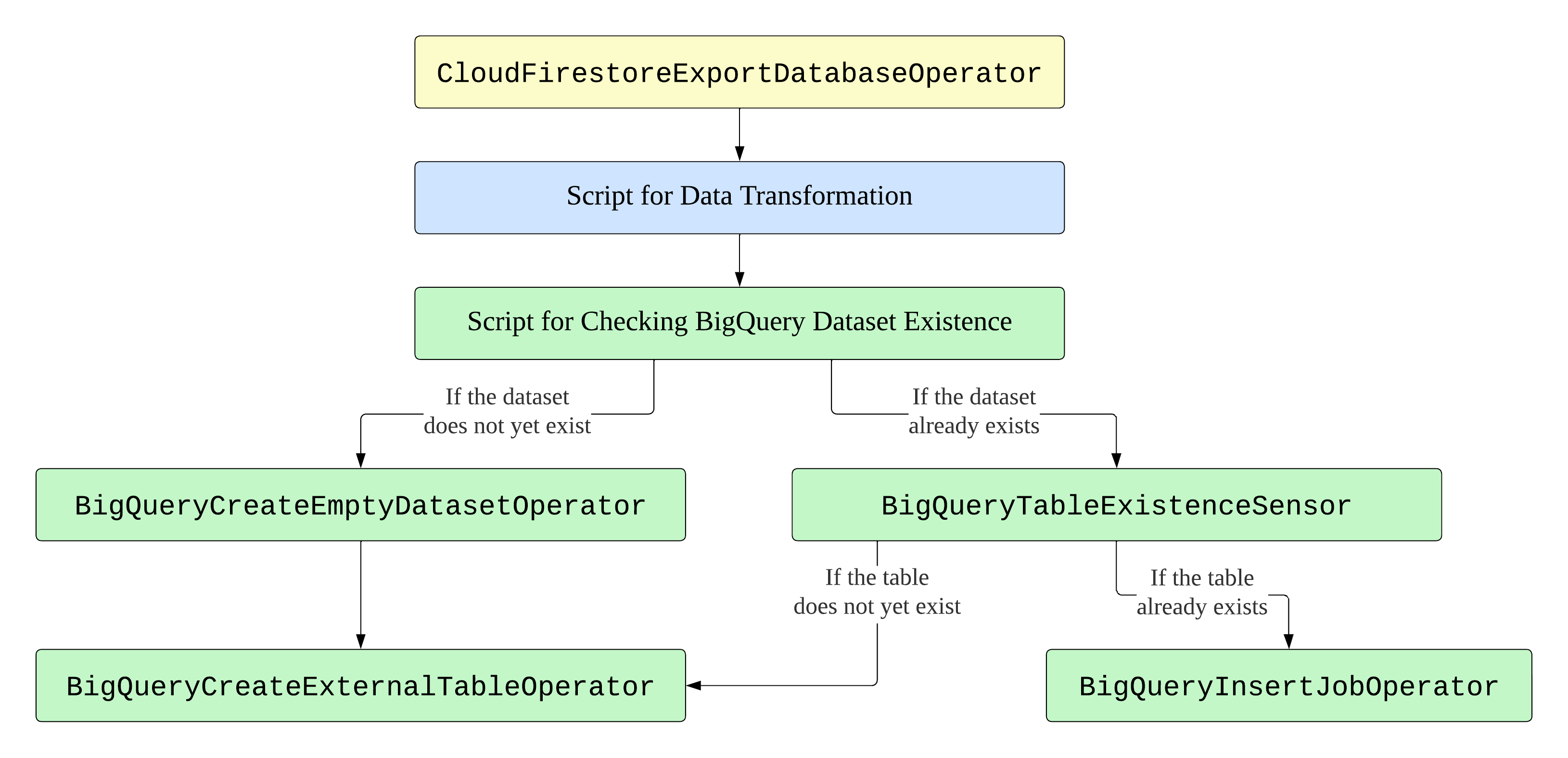}
    \caption{Directed acyclic graph representation of the ETL pipeline. \textmd{In the extraction stage (yellow), the documents storing the processed dialogues are fetched from Cloud Firestore. In the transformation stage (blue), a Python script modifies these documents to conform to the schema. Lastly, in the loading stage (green), they are stored as rows in the pertinent BigQuery tables.}}
    \label{etl}
\end{figure}

Following the star schema described in Section~\ref{star-schema-sec}, a relational database was created and stored in BigQuery \cite{18]bigquery}, a Google Cloud platform-as-a-service for data warehousing. An ETL pipeline was built for automating and scheduling the transfer of the processed dialogues in Cloud Firestore to the database in BigQuery. Figure~\ref{etl} presents the high-level directed acyclic graph representing this pipeline as implemented in Apache Airflow \cite{5]apache}, a workflow management platform for data engineering.

The script for the transformation stage has two primary functions. The first function maps the fields in the JSON document to their respective tables in the star schema. Its implementation is fairly straightforward since the prefix of the field names related to the review of systems is indicative of the pertinent table. The second function handles multi-valued fields and splits them into multiple fields following the logic discussed in Section~\ref{star-schema-sec}.

\subsection{Data Summary and Visualization}

The final stage in the process is creating a web-based dashboard for summarizing and visualizing health data from the children’s conversations with the chatbot. This dashboard was created using the visual analytics platform Tableau. In order to fetch the data, a live connection between the BigQuery database and the dashboard was established using Tableau’s BigQuery driver. 

\subsubsection{Dashboard Design}
\begin{figure*}[!t]
    \subfloat[Rolled Up]{%
    \includegraphics[width=0.25\linewidth]{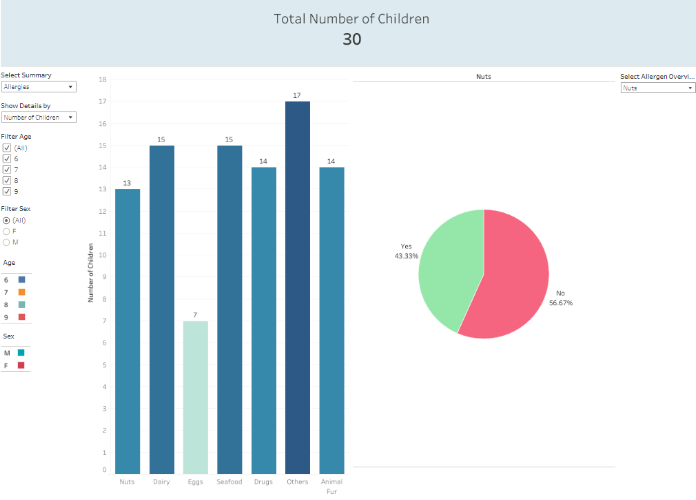} \label{6a}}
    \subfloat[Drilled Down by Age]{%
   \includegraphics[width=0.25\linewidth]{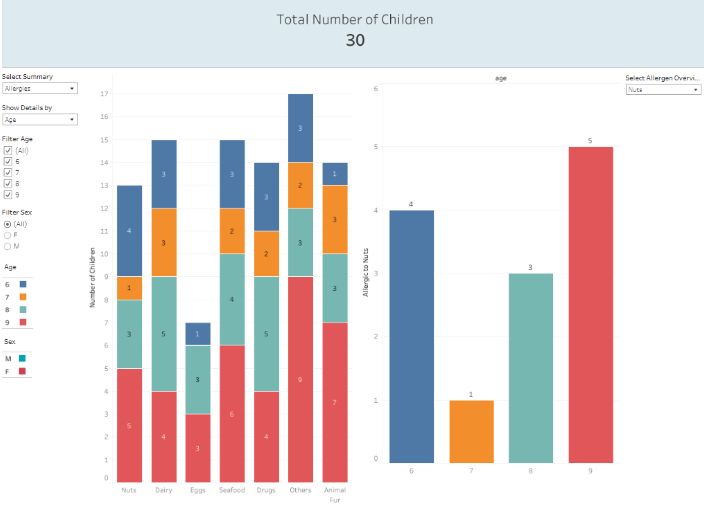} \label{6b}}
    \subfloat[Drilled Down by Sex]{%
   \includegraphics[width=0.25\linewidth]{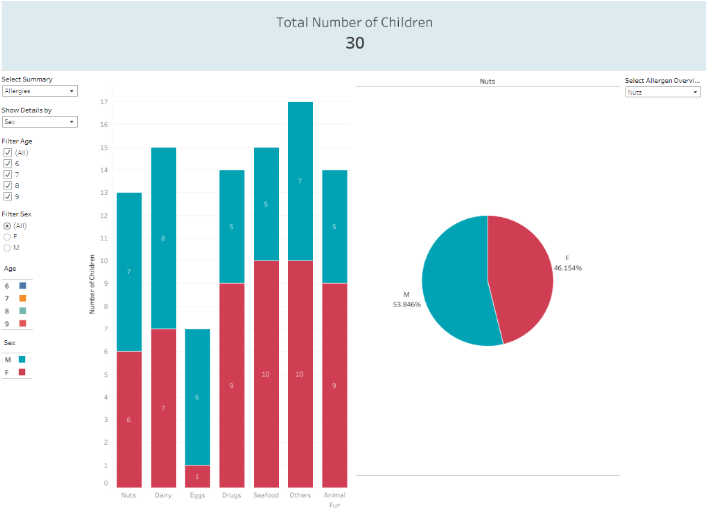} \label{6c}}
    \subfloat[Drilled Down by Age then Sex]{%
   \includegraphics[width=0.245\linewidth]{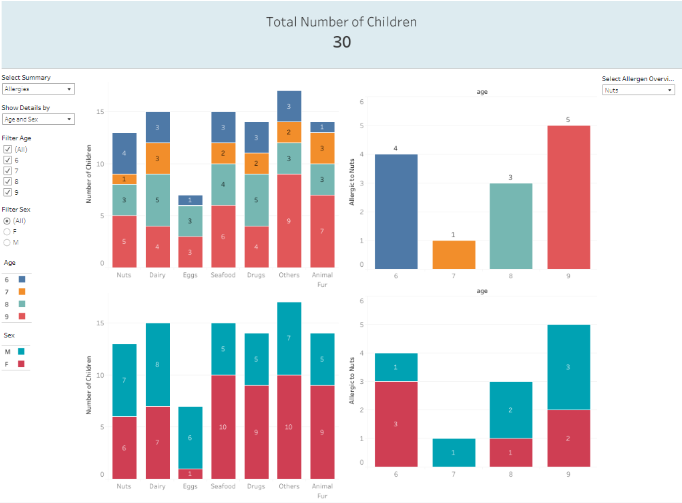} \label{6d}}
    \caption{Applying roll-up and drill-down on allergy data.  \textmd{Roll-up and drill-down operations can be executed via a dropdown on the left panel of the dashboard. As opposed to the aggregate statistics in Figure~\ref{6a}, the stacked bar graphs in Figure~\ref{6b} to Figure~\ref{6d} present data at a finer granularity. The graphs on the left side compare statistics across each of the common allergies, whereas those on the right focus on a particular allergy (selected via the dropdown on the right panel).}}
    \label{roll-up}
\end{figure*}

The dashboard consists of several charts corresponding to questions of interest in the adapted Pediatric Review of Systems. The design is intended to enable interactive and intuitive support for multiple OLAP operations. As such, it follows an underlying data hierarchy to allow users to change the granularity, i.e., perform roll-up and drill-down by demographic information (namely \textit{age} and \textit{sex}), as shown in Figure~\ref{roll-up}. 

\begin{figure}[!t]
    \subfloat[Filtered by Demographics]{%
    \includegraphics[width=0.5\linewidth]{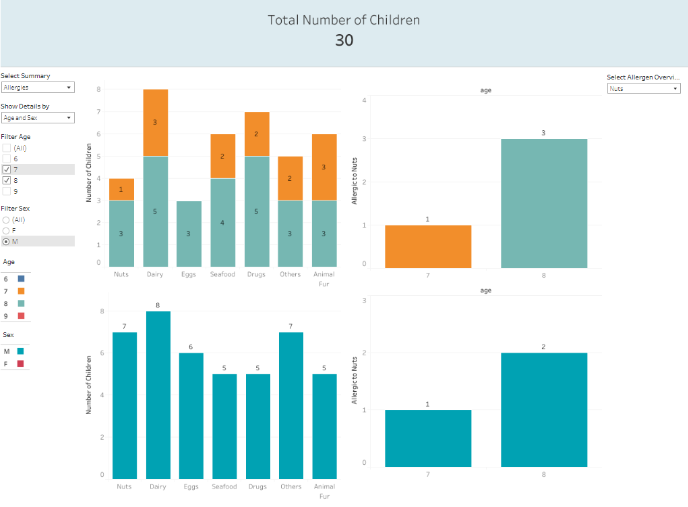} \label{7a}}
    \subfloat[Filtered by Response]{%
   \includegraphics[width=0.5\linewidth]{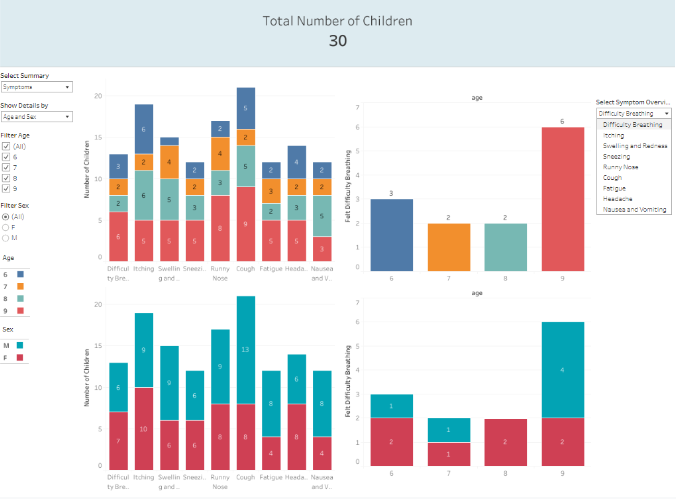} \label{7b}}
    \caption{Filtering allergy data. \textmd{To provide more focused insights, the drill-down operation in Figure~\ref{roll-up} can be complemented with slice and dice operations, which can be executed via the age and sex filters found on the left panel (Figure~\ref{7a}); the response filter dropdown on the right panel also dynamically changes depending on the selected summary visualization (Figure~\ref{7b}). The summary statistics for all the responses (left side of the dashboard) and the filtered charts (right side) are displayed vis-à-vis each other to allow users to compare and contextualize them without switching views.}}
    \label{filter}
\end{figure}

It also permits slicing and dicing by providing an option for users to dynamically filter the data based on demographic information and response, \rev{as shown in  Figure~\ref{filter}}. A dashboard with dynamic filters was used not only for compiling all the charts into one setting but also for presenting combinations of charts and adding interactivity. This ties to the intention of providing users with presentation flexibility, thus allowing them to understand the given health data more easily and positively facilitating their processing of the information \cite{56]yang2020}. 

In line with the common guidelines of public health dashboard design as mentioned in the work of Lechner and Fruhling \cite{32]lechner2014}, the dynamic filters also allow viewers to use a drill-down functionality to show more detailed visualizations from a more general summary, and present comparisons of data based on \textit{age} and \textit{sex}.

The data visualization consists primarily of three types of charts: \textit{(i)} pie charts for queries that admit only a single response (e.g., comparing the number of children who are allergic to nuts versus those who are not), \textit{(ii)} bar charts for those that admit multiple responses (e.g., comparing the number of children across each of the common allergies), and \textit{(iii)} stacked bar charts for drill-down queries (e.g., comparing the number of children across each of the common allergies and zeroing in on their age as well). 

\subsubsection{Performance Evaluation}
The performance of the dashboard was evaluated by measuring the loading time of different views. The tests were conducted on a machine with an AMD Ryzen\textsuperscript{TM} 7 5800H octa-core processor (3.20 GHz processor base frequency), 16.0 GB memory, 475 GB hard disk drive, and Microsoft Windows 10 (64-bit) operating system. To discard the effects of confounding variables, such as Internet connectivity and bandwidth, all the tests were conducted on the desktop version of Tableau, and the data were fetched from a live MySQL server hosted locally.

Aside from the initial loading time, the loading times of selected OLAP operations were also measured to provide a picture of the performance under varying numbers of stored sessions (dialogues) with the chatbot. The selected operations are as follows:   \textit{(i)} roll-up, \textit{(ii)} one-level drill-down (by age), \textit{(iii)} two-level drill-down (by age then sex), \textit{(iv)} single filter (by response), and \textit{(v)} double filter (by age and sex). Table~\ref{results} presents the results averaged over five trials. 

\begin{figure}[!t]
   \includegraphics[width=\linewidth]{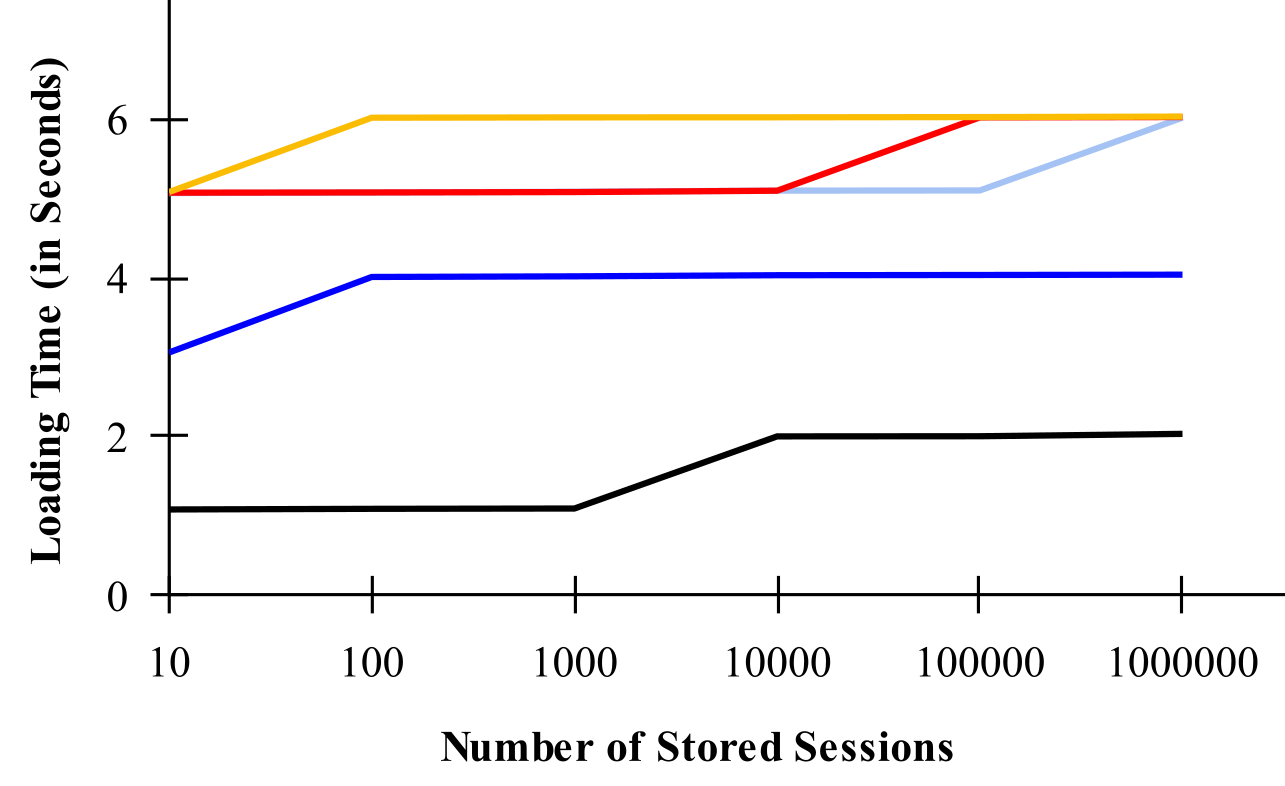} \label{8b}
    \caption{Linear-log plot of the loading times for selected OLAP operations. \textmd{The graph shows the loading times for roll-up
(sky blue), drill-down by age (red), drill-down by age and sex
(yellow), filter by response (black), and filter by age and sex
(blue).}}
    \label{results-graph}
\end{figure}

The initial loading time includes fetching data from the server, joining the tables, caching the data, and generating the first user-facing charts. Since a bulk of this time is spent communicating with the server and preparing the data for interactive actions, the design of the logical schema (Section~\ref{star-schema-sec}) is critical in this regard. Moreover, as seen in Figure~\ref{results-graph}, an increase in the number of stored entries by a factor of \(10^5\) increases the loading time for the display of roll-up, drill-down, and filter results by around only one second. 

\begin{table*}[t!]
  \caption{Loading times for the display of results of selected OLAP operations (in seconds)}
  \begin{tabular}{rrrrrrr}
    \toprule
    Num. of Stored & Initial Loading & Roll-Up & Drill-Down  & Drill-Down  & Filter  & Filter  \\
    Sessions & & & By Age & By Age then Sex &  By Response &  By Age and Sex \\
    \midrule
    10	& 16.089 & 5.063 & 5.068 & 5.076	& 1.080	& 3.056\\
    100	& 17.031 & 5.070	& 5.071	& 6.012	& 1.088	& 4.007\\
    1000 & 17.065 & 5.080 & 5.076 & 6.016 & 1.092 & 4.015\\
    10000 & 21.082 & 5.094 & 5.093 & 6.017 & 2.000 & 4.029\\
    100000 & 23.032 & 5.096 & 6.016 & 6.020 & 2.002 & 4.032\\
    1000000	 & 331.044 & 6.013 & 6.020 & 6.026 & 2.032 & 4.036\\
  \bottomrule
\end{tabular}
  \label{results}
\end{table*}

Reducing the overhead in the display of the results of these OLAP operations involves a trade-off between data caching and data freshness. This dashboard employs Tableau’s built-in drill-down, roll-up, and drill-down features, which operate on cached data instead of the more computationally expensive approach of communicating with the server for every OLAP query. These cached data are updated periodically, with the default being \(12\) hours \cite{45]datafreshness}. This trade-off was considered acceptable in the context of this project since the goal is to aggregate data over a period of time as opposed to generating real-time analytical reports.

\section{Conclusion}
Deriving insights from raw conversations necessitates taking into account the expressiveness of natural language and database considerations. This paper presented a four-stage \rev{process} for transforming unstructured dialogues into a structured data mart schema for visualization. First, the dialogues are processed by performing entity extraction and data aggregation using DialogFlowCX. Afterwards, the processed dialogues are stored as documents in Cloud Firestore. They are then transformed into a star schema for OLAP, with the relational database stored in BigQuery, and an ETL workflow built with Apache Airflow. Finally, an interactive web-based visualization for summarizing the data from conversations with the chatbot provides insights for health analytics. 

Evaluating the performance of this dashboard showed that increasing the number of stored sessions (dialogues) with the chatbot by a factor of \(10^5\) increased the loading time for the display of roll-up, drill-down, and filter results by around only one second. The analytics provided by this visualization may serve as a source of information on common sicknesses, symptoms, and health complaints of public school students in the Philippines. These present valuable insights for healthcare professionals, as well as those in charge of medical budgeting and intervention planning.

Future directions include improving the answer integrator module by expanding the entity synonym list with ontology-driven dictionaries and exploring multilingual word embeddings, \rev{as well as validating the current group of keywords and entity synonyms with domain experts}. Usability tests may also be conducted as a basis to improve the design and interactivity of subsequent iterations of the visualization dashboard.

\begin{acks}
  This work is part of a project funded by the Department of Science and Technology – Philippine Council for Industry, Energy, and Emerging Technology Research and Development (DOST-PCIEERD). 
\end{acks}

\bibliographystyle{ACM-Reference-Format}
\bibliography{References}

\end{document}